\input{aipcheck}

\documentclass[
    ,final            
  ]
  {aipproc}

\layoutstyle{8x11double}

\begin{document}

\title{Molecular Dynamics Simulation of Soft Grains: Malaria-Infected Red Blood Cells Motion within Obstructed 2-D Capillary Vessel}


\author{L. Haris}{
  address={Nuclear Physics and Biophysics Research Division, Faculty of Mathematics and Natural Sciences,
 Institut Teknologi Bandung, Jalan Ganesha 10, Bandung 40132, Indonesia
}
}

\author{S. N. Khotimah}{
  address={Nuclear Physics and Biophysics Research Division, Faculty of Mathematics and Natural Sciences,
 Institut Teknologi Bandung, Jalan Ganesha 10, Bandung 40132, Indonesia
}
}

\author{F. Haryanto}{
  address={Nuclear Physics and Biophysics Research Division, Faculty of Mathematics and Natural Sciences,
 Institut Teknologi Bandung, Jalan Ganesha 10, Bandung 40132, Indonesia
}
}

\author{S. Viridi}{
  address={Nuclear Physics and Biophysics Research Division, Faculty of Mathematics and Natural Sciences,
 Institut Teknologi Bandung, Jalan Ganesha 10, Bandung 40132, Indonesia
}
  ,altaddress={Computational Science, Faculty of Mathematics and Natural Sciences,
 Institut Teknologi Bandung, Jalan Ganesha 10, Bandung 40132, Indonesia
} 
}

\begin{abstract}
Molecular dynamics has been widely used to numerically solve equation of motion of classical many-particle system. It can be used to simulate many systems including biophysics, whose complexity level is determined by the involved elements. Based on this method, a numerical model had been constructed to mimic the behaviour of malaria-infected red blood cells within capillary vessel. The model was governed by three forces namely Coulomb force, normal force, and Stokes force. By utilizing two dimensional four-cells scheme, theoretical observation was carried out to test its capability. Although the parameters were chosen deliberately, all of the quantities were given arbitrary value. Despite this fact, the results were quite satisfactory. Combined with the previous results, it can be said that the proposed model were sufficient enough to mimic the malaria-infected red blood cells motion within obstructed capillary vessel.  
\end{abstract}

\maketitle


\section{Introduction}
Molecular dynamics has been widely used to numerically solve equation of motion of classical
many-particle system \cite{Rapaport:2004, Thijssen:2007}. It can be used to simulate many systems including biophysics, whose complexity level is determined by the involved elements \cite{Rapaport:2004}. Previously, this method has been numerously employed to model and to simulate malaria-infected red blood cells motion within capillary vessel \cite{Haris:2012, Haris:2012T, Haris:2013}. Throughout these simulations, the model undergo some alterations which were made to better mimic the biological system. Furthermore, some sort of rules have been established to avoid non-physical results \cite{Haris:2012T}. In this work, theoretical observation was carried out by using two dimensional four-cells scheme on our latest model. The goal was to ensure that the model was able to produce similar behaviours as it should be in the biological system.    

\section{Model of Red Blood Cells and Capillary Vessel}
There are three elements that are involved in the model; blood plasma, red blood cells, and capillary wall (a.k.a endothelial cell). Upon establishing a number of assumptions, the model consisting those elements was made in Cartesian coordinate as illustrated in Figure 1. All types of red-blood cells, both healthy and infected, are given spherical shape. Meanwhile, the capillary vessel is constructed as a set of rigid plates.

\begin{figure}
  \includegraphics[height=.3\textheight]{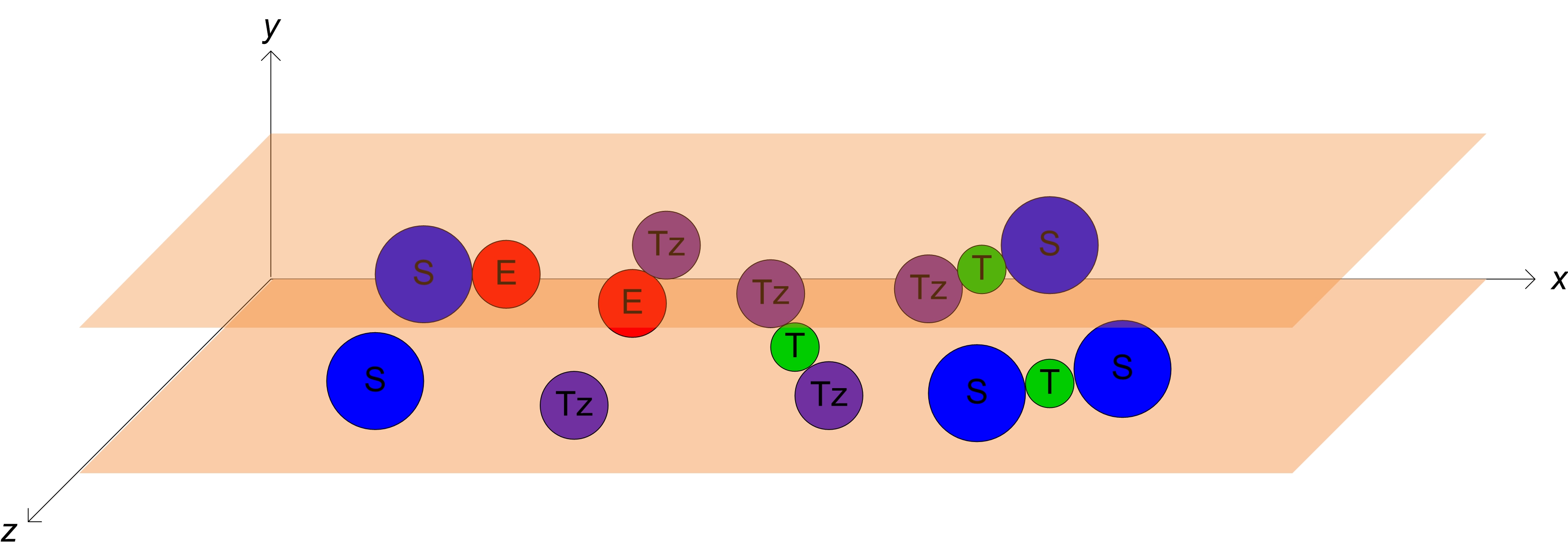}
  \caption{Model of red-blood cells and capillary vessel in Cartesian coordinate. All of the possible red-blood cells interactions are presented \cite{Kamolrat:1999}. E, T, Tz, and S are markers for red-blood cell, platelet, throphozoite, and schizont respectively.}
\end{figure}  

\section{Governing Equations}
Red-blood cells movement were governed by three forces; Coulomb force, repulsive force, and Stokes force. The first two forces are interactive forces between two red-blood cells. Since Coulomb force was defined as the attractive force between cells, repulsive force also need  to be defined to prevent two or more cells from collapsing to each other \cite{Shafer:1996}. Compared to the red-blood cells, the size of blood plasma molecules are considerably smaller. Thus, its flow can be considered as continuous one enabling the use of Stokes force.

There is an alteration made to the interactive forces formulation when dealing with capillary wall. The alteration was made on the vector unit that indicates the direction of the forces. The formulation of both the new vector unit and all of the three forces are provided in the following material \cite{Haris:2013Thesis}.
     
\section{Two Dimensional Four-Cells Scheme}
In this work, the problem was reduced to two dimensions in which four red-blood cells were involved; one healthy red-blood cells and three infected red-blood cells. In terms of the interactions between them, only rosetting and cytoadherence were considered in this scheme. Moreover, it was assumed that these clumps had occurred prior the simulation. The rosetting clump was given name as binary grain while two cytoadherence clumps were called obstacles. This scheme is illustrated in Figure 2.

The observation was done by varying several quantities such as initial binary grain orientation $(\theta_{0})$, initial binary grain height $(y_{0})$, and obstacles' tendency $(q)$, and blood plasma velocity $(v_{px})$. The value of these treatments are given in Table 1. Moreover, the initial properties of both binary grain and obstacles are provided within Table 2. Once varied, binary grain movement will be observed to obtain the clogging occurrence.

\begin{figure}
  \includegraphics[height=.3\textheight]{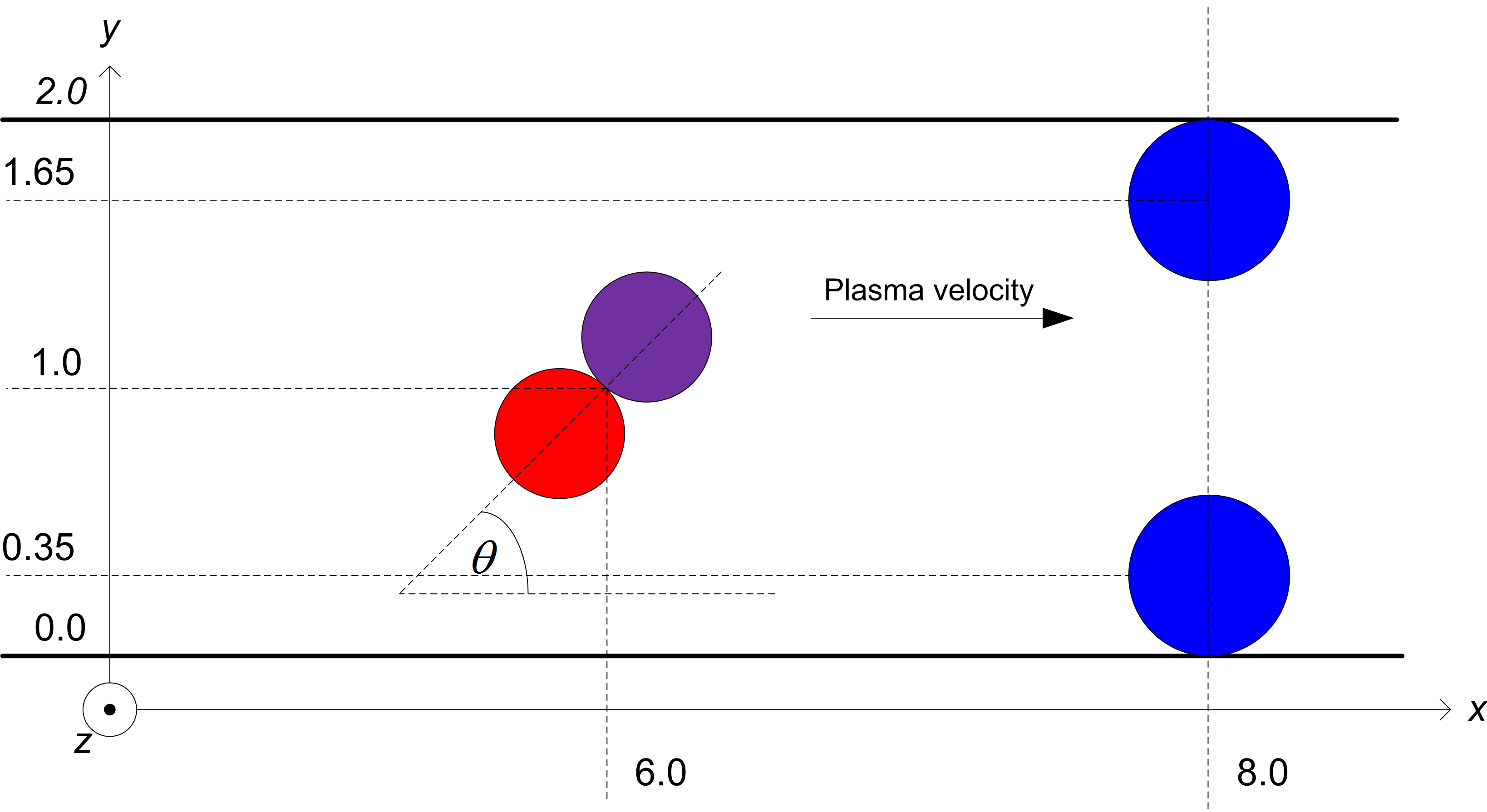}
  \caption{Four-cells scheme in two dimensions Cartesian coordinate. Red-blood cells, trophozoite, and schizonts are drawn as grains. They are recognizable by their colour; red, purple, and blue respectively. Two bold lines at $y=0.0$ and $y=2.0$ are capillary walls.}
\end{figure}  

\begin{table}
\begin{tabular}{c l c}
\hline
\tablehead{1}{c}{}{Quantities} & \tablehead{1}{c}{}{Range} & \tablehead{1}{c}{}{Interval} \\
\hline
$\theta_{0}$ & $0^{\circ}\ldots180^{\circ}$  & $10^{\circ}$\\
$y_{0}$ & $0.4\ldots1.6$ & $0.15$\\
$q$ & $0.0,-0.2$ & - \\
$v_{px}$ & $0.1\ldots16.3$ & $4.05$ \\
\hline
\end{tabular}
\caption{Treatments given to the proposed model in two-dimensional four-cells scheme.}
\end{table}

\begin{table}
\begin{tabular}{c c c c c}
\hline
& \tablehead{1}{c}{}{Trophozoite\\(Purple)} & \tablehead{1}{c}{}{Red-Blood Cell\\(Red)} & \tablehead{1}{c}{}{Upper Obstacle\\(Blue)} & \tablehead{1}{c}{}{Lower Obstacle\\(Blue)}\\
\hline
$\overrightarrow{r}_{0}$ & $\overrightarrow{r}_{\textrm{trophozoite}}$ & $\overrightarrow{r}_{\textrm{red blood cell}}$ & (8.0, 1.65, 0.0) & (8.0, 0.35, 0.0)\\
$\overrightarrow{v}_{0}$ & (0.0, 0.0, 0.0) & (0.0, 0.0, 0.0) & (0.0, 0.0, 0.0) & (0.0, 0.0, 0.0) \\
$q$ & -0.2 & 0.2 & $q$ & $q$ \\
$d$ & 0.4 & 0.4 & 0.7 & 0.7 \\
$m$ & 0.5 & 0.5 & 0.5 & 0.5 \\
\hline
\end{tabular}
\caption{Initial properties of binary grain and obstacles. $d$ and $m$ are diameter and mass respectively.}
\end{table}
  
\section{Simulation Procedure}
Improved Euler method was chosen in the molecular dynamics. For all cells we calculate,
\begin{equation}
\overrightarrow{v}_{i}\left(t_{k+1}\right)=\overrightarrow{v}_{i}\left(t_{k}\right)+\Delta t \overrightarrow{a}_{i}\left(t_{k}\right)
\end{equation}
\begin{equation}
\overrightarrow{r}_{i}\left(t_{k+1}\right)=\overrightarrow{r}_{i}\left(t_{k}\right)+\Delta t \overrightarrow{v}_{i}\left(t_{k}\right)
\end{equation}
where $k$ is the time index, and $\overrightarrow{a}$ was obtained through Newton's 2nd Law of Motion. The rest of the simulation parameters are provided within Table 3.

\begin{table}
\begin{tabular}{c c c c}
\hline
\multicolumn{2}{c}{\textbf{Grains-Planes}} & \multicolumn{2}{c}{\textbf{Grains-Grains}} \\ \hline
$k_{q}$ & 10.0 & $k_{q}$ & 10.0 \\
$k_{r}$ & $2\times 10^{5}$ & $k_{r}$ & $2\times 10^{5}$ \\
$k_{v}$ & 0.0 & $k_{v}$ & 5.0 \\ \hline
\multicolumn{2}{c}{$t$} & \multicolumn{2}{l}{1000} \\ 
\multicolumn{2}{c}{$\eta$} & \multicolumn{2}{l}{$1.2\times 10^{-3}$} \\
\multicolumn{2}{c}{$\Delta t$} & \multicolumn{2}{l}{$10^{-5}$} \\ \hline 
\end{tabular}
\caption{Treatments given to the proposed model in two-dimensional four-cells scheme.}
\end{table}

The simulation was run using the code written in C++ language, and was visualized using GLUT. The visualization was used together with the quantitative data to determine clogging occurrence.

\section{Results and Discussion}
Previously, it has been tested that the interactive forces worked well in modeling the interactions between healthy red-blood cells and the infected one \cite{Haris:2012, Haris:2012T}. Moreover, it was found that for charged obstacles, binary grain tends to rotate to a certain angle \cite{Haris:2013}. Now, using the parameters revealed within Table 1 up to Table 3, the binary grain movement was observed. 

\begin{figure}
  \includegraphics[height=.23\textheight]{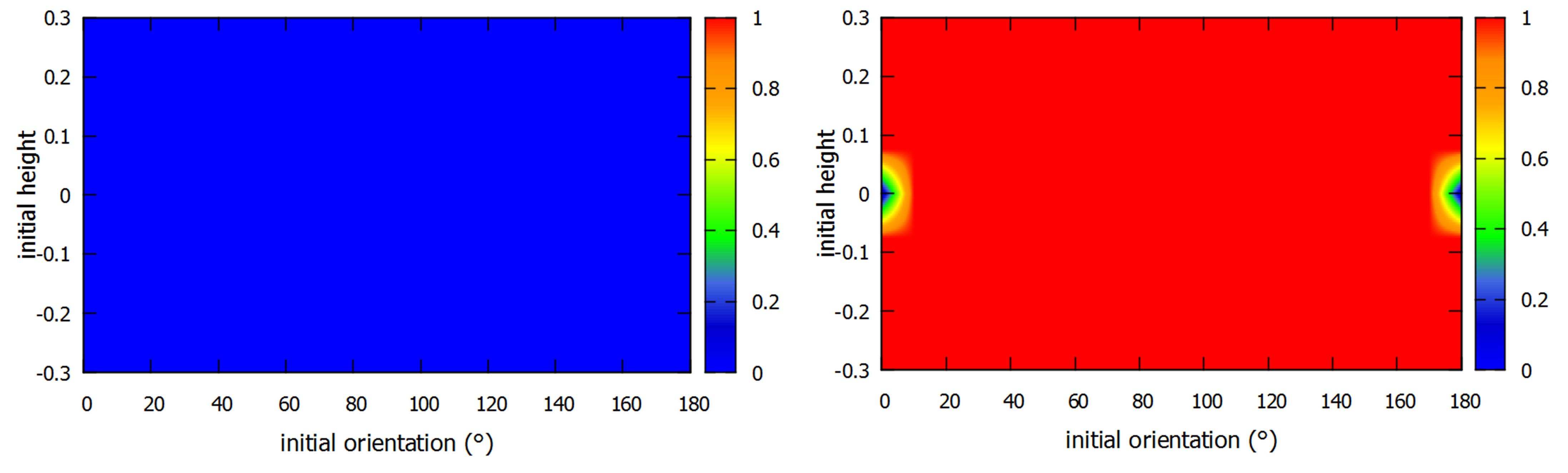}
  \caption{Clogging occurrence relative to binary grain's initial orientation and initial height for $q=0.0$ (left) and $q=-0.2$ (right).}
\end{figure}
\begin{figure}
  \includegraphics[height=.23\textheight]{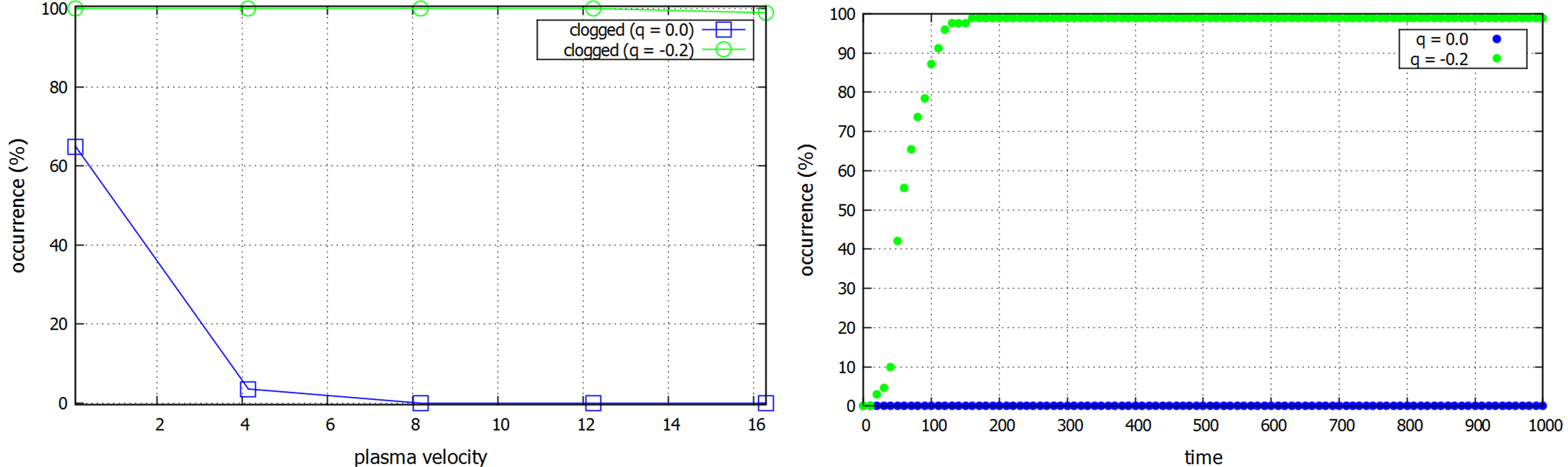}
  \caption{(Left) Capillary clogging occurrence due to the change of plasma velocity of the corresponding scheme for two different obstacles' tendency. (Right) The necessary time required by the binary grain to block the capillary vessel.}
\end{figure}

Figure 3 provides the contours of clogging occurrence relative to binary grain's initial angle $(\theta_{0})$ and initial height $(y_{0})$ for two different cases; $q=0.0$ and $q=-0.2$. It was found from these contours that if the obstacles lack the tendency to attract other blood cells $(q=0.0)$, the vessel will not be blocked. On the contrary, if the tendency is dominant $(q=-0.2)$, the vessel will surely be blocked except for some cases where $\theta_{0}=0^{\circ}$ and $\theta_{0}=180^{\circ}$. 

Figure 4 (left) is the graph of the clogging occurrence when there is a change in plasma velocity. The change may occur when one or more vessel are blocked. It is shown that the clogging occurrence decreases as plasma velocity increases. Meaning, when other vessels are blocked the clogging occurrence in another vessel will decrease. This decrement is significant when the obstacles' tendency to attract other red-blood cells are not dominant.

The last graph is the right side of Figure 4 which gives the information on the time needed for the vessel to be clogged. It was found that the model needs approximately 200 seconds to cause the clogging. This information may be used to cut the simulation time. However, in order to ensure the accuracy of the data, the simulation time is not cut. 
     
\section{Conclusion}
Although most of the parameters were determined deliberately, all of the quantities were given arbitrary value.Despite this fact, the resulting phenomena were quite satisfactory. Hence, the model can be said to be sufficient enough in mimicking the red blood cells movement within capillary vessel in the case of malaria.  

\begin{theacknowledgments}
Authors would like to extend their thanks to Riset Inovasi dan KK (RIK) ITB with contract no. 241/I.1.C01/PL/2013 for supporting this research so that it can be presented in the Symposium on Biomathematics 2013.
\end{theacknowledgments}

\bibliographystyle{aipproc}   
\bibliographystyle{aipprocl} 

\end{document}